%% file: paper.tex
\pdfoutput=1

\documentclass[runningheads]{llncs}
\usepackage{graphicx}
\DeclareGraphicsExtensions{.pdf}
\usepackage[T1]{fontenc}
\PassOptionsToPackage{hyphens}{url}\usepackage{hyperref}
\usepackage[per-mode=fraction]{siunitx}
\usepackage{amsmath}
\usepackage{amssymb}
\usepackage{listings}
\usepackage{here}
\usepackage{seqsplit}
\usepackage[bottom]{footmisc}
\makeatletter
\newcommand{\figcaption}[1]{\def\@captype{figure}\caption{#1}}
\newcommand{\tblcaption}[1]{\def\@captype{table}\caption{#1}}
\makeatother
\input{macros.tex}

\begin{document}
\title{Reinforcement Learning Testbed for Power-Consumption Optimization}
%
\author{Takao Moriyama
\and
Giovanni De Magistris
\and
Michiaki Tatsubori
\and \\
Tu-Hoa Pham
\and
Asim Munawar
\and
Ryuki Tachibana
}
\authorrunning{T. Moriyama et al.}
\institute{IBM Research -- Tokyo, Japan\\
\email{ \{moriyama,giovadem,mich,pham,asim,ryuki\}@jp.ibm.com}
}
\maketitle
\begin{abstract}
Common approaches to control a data-center cooling system rely on approximated system/environment models that are built upon the knowledge of mechanical cooling and electrical and thermal management.
These models are difficult to design and often lead to suboptimal or unstable performance.
In this paper, we show how deep reinforcement learning techniques can be used to control the cooling system of a simulated data center.
In contrast to common control algorithms, those based on reinforcement learning techniques can optimize a system's performance automatically without the need of explicit model knowledge.
Instead, only a reward signal needs to be designed.
We evaluated the proposed algorithm on the open source simulation platform EnergyPlus.
The experimental results indicate that we can achieve 22\% improvement compared to a model-based control algorithm built into the EnergyPlus.
To encourage the reproduction of our work as well as future research, we have also publicly released an open-source EnergyPlus wrapper interface \footnote{\url{https://github.com/IBM/rl-testbed-for-energyplus}} directly compatible with existing reinforcement learning frameworks.
\end{abstract}
\input{sections/introduction}
\input{sections/powerconsumption}
\input{sections/environment}
\input{sections/method}
\input{sections/results}
\input{sections/conclusion}
\bibliographystyle{splncs04}
\bibliography{bibs/energyplus}
\end{document}

%% file: macros.tex
\newcommand{\Astateset}{S}
\newcommand{\Astatevec}{\mathbf{s}}
\newcommand{\Aactionset}{A}
\newcommand{\Aactionvec}{\mathbf{a}}
\newcommand{\Arewardval}{r}
\newcommand{\Atransitionprobability}{P}
\newcommand{\Arewardfunction}{R}
\newcommand{\Ainitialstateprobability}{\rho_{0}}
\newcommand{\Adiscountfactor}{\gamma}
\newcommand{\Apolicy}{\pi}
\newcommand{\Adiscountedexpectedreturn}{\eta}
\newcommand{\Aparentheses}[1]{{\left(#1\right)}}
\newcommand{\Abrackets}[1]{{\left[#1\right]}}
\newcommand{\Aexpectation}{\mathop{\mathbb{E}}}
\newcommand{\Astateactiontrajectory}{\tau}

\newcommand{\Aneuralnetworkparameters}{\theta}
\newcommand{\Amean}{\mu}
\newcommand{\Astd}{\sigma}

%% file: sections/introduction.tex
\section{Introduction}
\label{sec:introduction}
Data centers worldwide consume around 3\% of the total electricity consumed globally.
This electricity consumption adds up to roughly 90 billion kilowatt-hours annually in the US\footnote{\url{https://www.forbes.com/sites/forbestechcouncil/2017/12/15/why-energy-is-a-big-and-rapidly-growing-problem-for-data-centers}}; thus, it is one of the largest issues regarding data centers.
The report of the US State Department of Energy identified that the effective energy saving policies reduce the energy consumption of data centers (620 billion kilowatt-hours will be saved between 2010 and 2020).
The design of hardware and software components is a key factor in saving energy.
In this study, we focused on improving the built-in control algorithm (hereafter, 'controller'), of one of the major sources of energy consumption, i.e., the cooling system~\cite{Singapore:2015:Benchmark}, by using recent advances in deep reinforcement learning (DRL).

A cooling system is composed of multiple components controlled by adjusting the desired values of different control temperatures called setpoints.
Common approaches to control them are based on approximate models of the system~\cite{Sun:else2005:HVACPlant}.
These models are sometimes inaccurate or too complex to calculate.
Therefore, a data-driven approach has recently been considered as an interesting alternative.
In such cases, the control policy is learned using the data acquired from the system without using any explicit system model.
We discuss the advantages of using this approach compared to the classical model-based approach.
We propose a controller for controlling the cooling system using a state-of-the-art DRL technique.
By designing a reward function considering power consumption and reference temperature ranges, our DRL controller outperforms the built-in controller by 22\% in terms of power consumption while maintaining a similar temperature range.
Compared to previous related studies~\cite{arXiv:Li:2017}\cite{Wei:DAC2017:DRLforHVAC}, we provide a DRL open source environment for EnergyPlus~\cite{Crawley2001:EnergyPlus} following the OpenAI Gym \footnote{\url{https://github.com/openai/gym}} interface.
This allows designers to easily test their algorithms and control strategies as well as using current algorithms based on the interface.

The paper is structured as follows.
In Section~\ref{sec:simulation}, we formally define the problem.
In Section~\ref{sec:environment}, we explain the open-source environment we built and released to train DRL controllers for energy optimization.
In Section~\ref{sec:method}, we describe our DRL controller and resolution method.
In Section~\ref{sec:results}, we present simulation results on different control configurations (e.g., classical vs. DRL) and environments (e.g., weather settings).
We conclude the paper in Section~\ref{sec:conclusion} with directions for future works.

%% file: sections/powerconsumption.tex
\section{Control of Heating, Ventilation, and Air-Conditioning System}
\label{sec:simulation}
The heat, ventilation, and air conditioning (HVAC) controller optimizes the data-center power consumption under operation constraints (e.g., maintaining temperatures in specified ranges for employees).
The design of this controller is not simple due to the complexity of the HVAC system (e.g., fluid dynamics and thermodynamics) and dependence of many external factors (e.g., external weather conditions).

\begin{figure}[t]
\centering
\includegraphics[scale=0.6, bb=0 45 579 485]{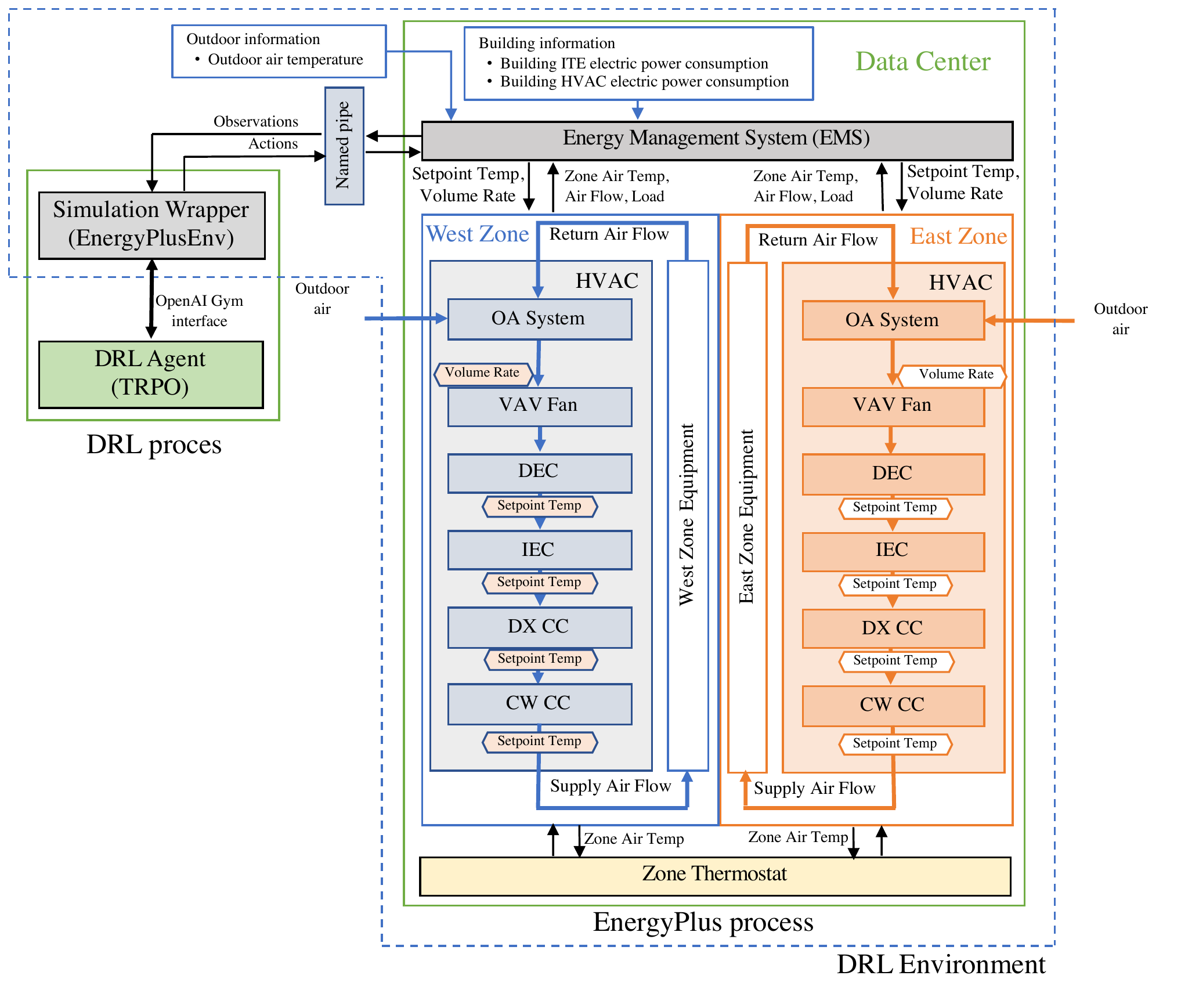}
\caption{Control of each zone of data center} 
\label{fig_rlenv}
\end{figure}

Figure~\ref{fig_rlenv} shows a model of a data center, which consists of two zones (server rooms).
Each zone has a dedicated HVAC system, which is connected through a ``supply air'' duct and ``return air'' duct to exchange heat.
Each HVAC system is composed of several components connected sequentially as follows:
\begin{itemize}
\item outdoor air system (OA System) exchanges zone and outdoor air flow
\item variable volume fan (VAV Fan) adjusts the air flow rate to meet the zone target temperature
\item direct evaporative cooler (DEC) lowers the air temperature using latent heat from the evaporation of water
\item indirect evaporative cooler (IEC) lowers the air temperature without adding humidity to the air
\item direct expansion cooling coil (DX CC) cools air by passing the condensed refrigerant through a heat exchanger
\item chilled water cooling coil (CW CC) cools air using chilled water
\end{itemize}
The target temperature for the DEC, IEC, DX CC, and CW CC are specified by a single common temperature, called the setpoint temperature, for each zone.
Using model-based approaches, the setpoint manager calculates the setpoint temperatures based on a built-in model for system dynamics and environmental information, system loads, etc.
The air volume supplied to each zone is also adjusted by the VAV Fan.
To apply DRL to the data-center model, we replace the model-based controller (setpoint manager and VAV Fan controller) with our DRL-based controller.

%% file: sections/environment.tex
\section{Reinforcement Learning Testbed for EnergyPlus}
\label{sec:environment}
\lstset{ 
  basicstyle=\scriptsize\ttfamily,
  frame=single,
  framexleftmargin=-2mm,
  framexrightmargin=-2mm
}
\subsection{Simulation Wrapper}
\label{sec:environement:wrapper}
Figure~\ref{fig_rlenv} illustrates our simulation system setup.
Our DRL controller is composed of two scripts: DRL-based agent and energy management system (EMS).
We use the OpenAI Baselines for open-source implementation of the trust-region policy optimization (TRPO) algorithm~\cite{code:baselines:2017} as the agent.
We explain the details of this algorithm in Section~\ref{sec:method}.

We define the simulation wrapper (EnergyPlusEnv) shown in Figure~\ref{fig_rlenv} to manage the interaction between the learning agent and EnergyPlus simulation.
We create this wrapper following the OpenAI Gym interface, which consists of the following two major APIs:
\begin{itemize}
\item \textbf{Env.reset()}: Restart the simulation process
\item \textbf{Env.step()}: Proceed one simulation timestep
\end{itemize}
During the start of the training, a new instance of EnergyPlusEnv is created, which then creates a pair of named pipes: one for sending information from EnergyPlusEnv to EnergyPlus and the other for the opposite direction.
When Env.reset() is called by the agent, EnergyPlusEnv spawns a new EnergyPlus process with the information about the name of the pipes, building model file (input data file, IDF), and weather file (EnergyPlus weather file, epw).
EnergyPlusEnv waits until the first observation is returned from the EnergyPlus simulation.
Env.step() sends the action to EnergyPlus and waits until observation is returned.
EnergyPlusEnv computes the reward and sends it along with the observations to the agent.
The EMS script in Figure~\ref{fig_rlenv} receives action information from EnergyPlusEnv and sets it to corresponding variables in the EnergyPlus simulation process.
It also gathers information from the EnergyPlus process and sends it to EnergyPlusEnv as observations.

The communication protocol between EnergyPlusEnv and the simulation consists of sending a series of floating-point values of action to the EnergyPlus simulation.
Similarly, we receive a series of floating-point values as observation back from the EnergyPlus simulation.
Therefore, we do not need to encode any command code but just send (receive) the amount of data, then the sequence of data follows for both directions.

\subsection{Extending Built-in Energy Management System}
\label{sec:environment:extendingems}
In this section, we explain how to hook into the process of EnergyPlus.
EnergyPlus provides two frameworks to extend its functionalities:
\begin{itemize}
\item Building Controls Virtual Test Bed (BCVTB): software framework for co-simulation, which allows importing of the functional mockup unit (FMU) from other systems, such as MATLAB/Simulink, to EnergyPlus and exporting EnergyPlus as an FMU to other simulation systems.
It is intended for building a large-scale and distributed simulation system including EnergyPlus as a component.
\item EMS: a high-level control method available in EnergyPlus.
The EMS uses a small scripting language called EnergyPlus Runtime Language (Erl) to access a variety of sensor data, make decisions based on these data, and select the control actions for actuators.
\end{itemize}

While the BCVTB is designed to connect two or more simulation systems, the EMS is designed to extend EnergyPlus capability inside it but does not have the capability to connect to outside the EnergyPlus system.
We used the EMS to extend EnergyPlus because of its simplicity and added two built-in functions to Erl to allow EMS scripts to communicate with external entities:
\begin{itemize}
\item \textbf{@ExtCtrlObs}: send observation vector to EnergyPlusEnv.
\item \textbf{@ExtCtrlAct}: receive action vector from EnergyPlusEnv.
\end{itemize}
We selected a two-zone data-center model (``2ZoneDataCenterHVAC\_wEconomizer.idf'') because it has combination of different types of HVAC components (DEC, IEC, single speed DX CC, and CW CC).
\textbf{@ExtCtrlObs} and \textbf{@ExtCtrlAct} are used in Erl script as follows:
\begin{lstlisting}
  1:  EnergyManagementSystem:Program,
  2:    ExtCtrlBasedSetpointManager,
  3:    SET tmp = @ExtCtrlObs 1 OutdoorTemp,
  4:    SET tmp = @ExtCtrlObs 2 WestZoneTemp,
  5:    SET tmp = @ExtCtrlObs 3 EastZoneTemp,
  6:    SET tmp = @ExtCtrlObs 4 Whole_Building_Power,
  7:    SET tmp = @ExtCtrlObs 5 IT_Equip_Power,
  8:    SET tmp = @ExtCtrlObs 6 Whole_HVAC_Power,
  9:    SET tmp = @ExtCtrlAct 0 6,
 10:    SET WestZoneDECOutletNode_setpoint = @ExtCtrlAct 1,
 11:    SET WestZoneIECOutletNode_setpoint = @ExtCtrlAct 1,
 12:    SET WestZoneCCoilAirOutletNode_setpoint = @ExtCtrlAct 1,
 13:    SET WestAirLoopOutletNode_setpoint = @ExtCtrlAct 1,
 14:    SET EastZoneDECOutletNode_setpoint = @ExtCtrlAct 2,
 15:    SET EastZoneIECOutletNode_setpoint = @ExtCtrlAct 2,
 16:    SET EastZoneCCoilAirOutletNode_setpoint = @ExtCtrlAct 2,
 17:    SET EastAirLoopOutletNode_setpoint = @ExtCtrlAct 2,
 18:    SET WestZoneSupplyFan_FanAirMassFlowRate = @ExtCtrlAct 3,
 19:    SET EastZoneSupplyFan_FanAirMassFlowRate = @ExtCtrlAct 4;
\end{lstlisting}
\textbf{@ExtCtrlObs} specifies one of the elements of a state vector in turn (lines 3-8).
It is a function with two parameters.
The first parameter specifies the index in the state vector for the specified value in the second parameter.
The index starts from 1.
The specified value is stored in the internal buffer.

If \textbf{@ExtCtrlAct} is called with 0 as the first parameter (line 9), it sends the state vector with a length specified by the second parameter to EnergyPlusEnv through one of the named pipes then waits until an action vector is sent back from EnergyPlusEnv through another named pipe.
The received action vector is stored into the internal buffer.
The non-zero value for the first parameter of \textbf{@ExtCtrlAct} is treated as an index for an element in the internal buffer, and the specified element is returned (lines 10-19).
The value returned from \textbf{@ExtCtrlAct} is assigned to the actuator in the left side of the assignment statement.
As described in the code above, we have four elements in the action vector.
The first and second elements specify the setpoint temperature for West and East Zones, respectively.
The other two elements are used to set the air flow rate for West and East Zones.

By defining an ``EnergyManagementSystem:Sensor'' object as a system variable, it becomes accessible from EMS scripts.
\begin{lstlisting}
  EnergyManagementSystem:Sensor,
    WestZoneTemp,              !- Name
    WEST ZONE,                 !- Output:Variable or Output:Meter Index Key Name
    Zone Mean Air Temperature; !- Output:Variable or Output:Meter Name
\end{lstlisting}

However, by defining ``EnergyManagementSystem:Actuator'' for a control variable, it becomes controllable from EMS scripts as follows:

\begin{lstlisting}
  EnergyManagementSystem:Actuator,
    WestZoneDECOutletNode_setpoint, !- Name
    West Zone DEC Outlet Node,      !- Actuated Component Unique Name
    System Node Setpoint,           !- Actuated Component Type
    Temperature Setpoint;           !- Actuated Component Control Type
\end{lstlisting}
{\sloppy
We added three new sensors: OutDoorTemp, WestZoneTemp, and EastZoneTemp in addition to existing sensors regarding measuring electric demand power: IT\_Equip\_Power, Whole\_Building\_Power, and Whole\_HVAC\_Power.
We also defined ten actuators: WestZoneDECOutletNode\_setpoint,
WestZoneIECOutletNode\_setpoint,
WestZoneCCoilAirOutletNode\_setpoint,
WestAirLoopOutletNode\_setpoint,
EastZoneDECOutletNode\_setpoint,
EastZoneIECOutletNode\_setpoint,
EastZoneCCoilAirOutletNode\_setpoint,
EastAirLoopOutletNode\_setpoint,
WestZoneSupplyFan\_FanAirMassFlowRate,
and EastZoneSupplyFan\_FanAirMassFlowRate.

}

\subsection{Replacing Existing Controller with Agent-Based Controller}
\label{sec:environment:replaceexistingcontrol}
To control the temperature setting, we need to do the following two actions:
\begin{itemize}
\item disabling the existing controller in EnergyPlus
\item reporting the current state (observation) in EnergyPlus to the agent system, receiving the next actions from an agent, and reflecting these actions to the EnergyPlus simulation, as defined in the OpenAI Gym interface.
\end{itemize}
The temperature in the original model is controlled using the following ``ZoneControl:Thermostat'' objects for each zone.
\begin{lstlisting}
  ZoneControl:Thermostat,
    West Zone Thermostat,             !- Name
    West Zone,                        !- Zone or ZoneList Name
    Zone Control Type Sched,          !- Control Type Schedule Name
    ThermostatSetpoint:DualSetpoint,  !- Control 1 Object Type
    Temperature Setpoints;            !- Control 1 Name
\end{lstlisting}

``ZoneControl:Thermostat'' can operate with various control types in a predefined schedule.
These controls include single heating setpoint, single cooling setpoint, single heating/cooling setpoint, and dual setpoint (heating and cooling) with deadband.
In this case, only the dual setpoint is used with a constant high-limit temperature (\SI{23.0}{\celsius}) and low-limit temperature (\SI{20.0}{\celsius}).

The target nodes to which setpoint temperatures are controlled are defined by the ``SetpointManager:SingleZone:Cooling'' object as follows.
Note that the comments are indicated with exclamation marks.

\begin{lstlisting}
 !  SetpointManager:SingleZone:Cooling,
 !    System Setpoint manager, !- Name
 !    Temperature,             !- Control Variable
 !    -99.0,                   !- Minimum Supply Air Temperature {C}
 !    99.0,                    !- Maximum Supply Air Temperature {C}
 !    West Zone,               !- Control Zone Name
 !    West Zone Air Node,      !- Zone Node Name
 !    West Zone Inlets,        !- Zone Inlet Node Name
 !    West Sys Setpoint nodes; !- Setpoint Node or NodeList Name
\end{lstlisting}

``West Sys Setpoint nodes'' and ``East Sys Setpoint nodes'' define a list of actual node names.
To disable the existing controller, we disabled SetpointManager:SingleZone:Cooling in the simulation model by simply commenting out, as shown above.
We simply override the ``fan air mass flow rate'' of the VAV Fan through actuators to control the air flow rate.

\subsection{How to Hook into Simulation Loop in EnergyPlus}
\label{sec:environment:howtohook}
EMS provides the ``calling point'' mechanism, which allows EMS users to invoke a specified EMS procedure in a specified point.
These points are called ``calling points''.
There are 14 calling points defined in the EMS.
We used ``AfterPredictorAfterHVACManagers'', which activates associated procedures after the predictor, and the traditional HVAC managers are called.
By registering the ExtCtrlBasedSetpointManager procedure we discussed in Section~\ref{sec:environment:extendingems} with ``AfterPredictorAfterHVACManagers'', we can control the setpoint temperatures of each zone.

\begin{lstlisting}
  EnergyManagementSystem:ProgramCallingManager,
    ExtCtrl-Based Setpoint Manager,       !- Name
    AfterPredictorAfterHVACManagers,      !- EnergyPlus Model Calling Point
    ExtCtrlBasedSetpointManager;          !- Program Name 1
\end{lstlisting}

\subsection{Application to Other Simulation Models}
In previous sections, we used the two-zone data-center model ``2ZoneDataCenterHVAC\_wEconomizer.idf'' to explain our changes with respect to the original file.
If you want to use a different data-center model, the following steps are necessary.

\begin{enumerate}
\item Select the building model (IDF file) you want to control using a DRL-based agent.
\item Determine which setpoint (control) you want to replace by using the DRL agent.
  Inhibit it in the IDF file, as described in Section~\ref{sec:environment:replaceexistingcontrol}
\item Determine the list of required sensors (observations) and actuators (actions).
  Add them by defining ``EnergyManagementSystem:Sensor'' and ``EnergyManagementSystem:Actuator'' objects in the IDF file (Section~\ref{sec:environment:extendingems}).
\item Write an Erl program in the IDF file to exchange actions and observation with the EnergyPlusEnv wrapper (Section~\ref{sec:environment:extendingems}).
\item Hook the Erl program into the EnergyPlus simulation loop by defining the ``EnergyManagementSystem:ProgramCallingManager'' object in the IDF file (Section~\ref{sec:environment:howtohook}).
\end{enumerate}

You also need to develop a model-dependent part of the simulation wrapper for your simulation model.
In our case, it is implemented using a Python code called the \lstinline{energyplus_model_2ZoneDataCenterHVAC_wEconomizer_FanControl.py} file, which implements the following methods:
\begin{itemize}
\item EnergyPlusModel.setup\_spaces(): Define action and observation spaces
\item EnergyPlusModel.compute\_reward(): Compute reward value from current states and actions
\item EnergyPlusModel.read\_episode(), EnergyPlusModel.plot\_episode(): Read result from CSV file generated by EnergyPlus and plot it on the screen. This is for visualization purpose.
\end{itemize}

%% file: sections/method.tex
\section{Reference Reinforcement Learning Implementation}
\label{sec:method}

\subsection{Model-Free Reinforcement Learning}
We consider the data-center control problem as an infinite-horizon discounted Markov decision process (MDP).
Such MDPs are characterized by
\begin{itemize}
    \item $\Astateset$ a set of states
        (e.g., all possible temperatures inside and outside a building),
    \item $\Aactionset$ a set of possible actions to control the system
        (e.g., cooling commands),
    \item $\Atransitionprobability: \Astateset \times \Aactionset \times \Astateset \rightarrow [0,1]$
        the state-action-state transition probability distribution
        (e.g., to go from a given temperature to another with a given command)
    \item $\Arewardfunction: \Astateset \times \Aactionset \times \Astateset \rightarrow \mathbb{R}$
        a reward function corresponding to such transitions
    \item $\Ainitialstateprobability: \Astateset \rightarrow [0,1]$
        the initial state probability distribution,
    \item $\Adiscountfactor \in [0, 1)$ a discount factor
        penalizing future reward expectations.
\end{itemize}
We are then interested in computing a stochastic policy
$\Apolicy: \Astateset \times \Aactionset \rightarrow [0, 1]$
that maximizes the discounted expected return
$\Adiscountedexpectedreturn\Aparentheses{\Apolicy}$:
\begin{align}
    \label{eq:discountedexpectedreturn}
    \Adiscountedexpectedreturn \Aparentheses{ \Apolicy } = 
    \Aexpectation\limits_{\tau}\Abrackets{
        \sum\limits_{i=0}^{\infty} {
            \Adiscountfactor^i \Arewardfunction\Aparentheses{
                \Astatevec_i, \Aactionvec_i, \Astatevec_{i+1}
            }
        }
    }%
    ,
\end{align}
with $\Astateactiontrajectory = \Aparentheses{\Astatevec_0, \Aactionvec_0, \Astatevec_1, \Aactionvec_1, \dots}$
a sequence of states and actions,
where the initial state $\Astatevec_0$ is initialized following $\Ainitialstateprobability$,
each action $\Aactionvec_i$ is sampled following
$\Apolicy\Aparentheses{\cdot | \Astatevec_i}$
the control policy given the current state $\Astatevec_i$,
leading to a new state $\Astatevec_{i+1}$
following the transition function
$\Atransitionprobability\Aparentheses{\cdot | \Astatevec_i, \Aactionvec_i}$.
Given Eq.~\eqref{eq:discountedexpectedreturn},
we seek to maximize a \emph{reward expectation over multiple steps}
rather than a \emph{direct one-step reward}
(which we could approach by, e.g., greedy search).
Thus, reinforcement learning approaches have to deal with major challenges,
including the fact that a reward at a given step is not necessarily associated to
a single action but possibly a sequence thereof (the credit assignment problem)
and the fact that the agent has to balance between exploiting known rewards
and exploring new strategies that can lead to higher but can also lower rewards
(the exploitation-exploration dilemma).
This is particularly true in the case of data-center optimization
since heat distribution occurs over time
(e.g., a given command may not impact sensor readings instantaneously)
and can depend on non-controllable parameters
(e.g., weather conditions).

DRL approaches have recently provided considerable results on such problems, where the relationship between states and optimal actions is difficult to model formally,
e.g.,
playing Atari video games directly from pixels
using convolutional neural networks (CNNs)
rather than handcrafted features~\cite{nature:mnih:2015},
or learning complex robotic tasks in both simulated~\cite{arxiv:heess:2017}
and real environments~\cite{iros:inoue:2017}.
In DRL, control policies $\Apolicy_{\Aneuralnetworkparameters}$
are typically represented by deep neural networks
parameterized by vector of parameters $\Aneuralnetworkparameters$
(e.g., neuron weights and biases).
In our study,
we used the TRPO algorithm~\cite{icml:schulman:2015},
in which
a multilayer perceptron (MLP) predicts
as outputs
the mean $\Amean$ and standard deviation $\Astd$ of a Gaussian distribution,
taking directly as input
a state vector $\Astatevec_i$.
Given a policy $\Apolicy$,
we denote the state-action value function as $Q_\pi$,
the value function as $V_\pi$,
and the advantage function as $A_\pi$:
\begin{align}
    Q_{\pi}(\Astatevec_i, \Aactionvec_i) &= \mathop{\mathbb{E}}\limits_{\Astatevec_{i+1}, \Aactionvec_{i+1}, \dots}\left[
    \sum\limits_{l=0}^{\infty}\gamma^l R(\Astatevec_{i+l}, \Aactionvec_{i+l}, \Astatevec_{i+l+1})
\right], \\
 V_{\pi}(\Astatevec_i) &= \mathop{\mathbb{E}}\limits_{\Aactionvec_i, \Astatevec_{i+1}, \dots}\left[
    \sum\limits_{l=0}^{\infty}\gamma^l R(\Astatevec_{i+l}, \Aactionvec_{i+l}, \Astatevec_{i+l+1})
\right], \\
A_{\pi}(\Astatevec, \Aactionvec) &= Q_{\pi}(\Astatevec, \Aactionvec) - V_{\pi}(\Astatevec).
\end{align}
The parameters $\Aneuralnetworkparameters$
of the neural network policy $\Apolicy_{\Aneuralnetworkparameters}$
are then iteratively refined by
collecting state-action-reward trajectories in the environment
and
solving the following problem:
\begin{align} \label{eq:trpo}
    &\Aneuralnetworkparameters_{k+1} = \mathop{\text{argmax}}\limits_{\Aneuralnetworkparameters}
    \mathop{\mathbb{E}}\limits_{\Astatevec \sim \rho_{\Aneuralnetworkparameters_k}, \Aactionvec \sim \pi_{\Aneuralnetworkparameters_k}} \left[
        \frac{\pi_{\Aneuralnetworkparameters}( \Aactionvec | \Astatevec )}{\pi_{\Aneuralnetworkparameters_k}(\Aactionvec|\Astatevec)}A_{\pi_{\Aneuralnetworkparameters_k}}(\Astatevec, \Aactionvec)
    \right]%
    ,
\end{align}
with additional constraints on parameter variation
to facilitate convergence~\cite{icml:schulman:2015}.
In particular, we use the OpenAI Baselines open-source implementation
of the TRPO algorithm~\cite{code:baselines:2017}.
In our study, we trained DRL controllers in a purely model-free fashion,
i.e.,
no specific knowledge on the data-center operation was pre-encoded in the neural networks
or training process,
instead using only abstract state and action vectors as inputs and outputs.
Note that models can be used to accelerate training when
available~\cite{icra:pham:2018},
and the quality of the resulting policy ultimately depends on the
quality of the models.
Domain adaptation and transfer learning~\cite{iros:tobin:2017,icip:inoue:2018}
constitute important steps towards the application of such model-based DRL techniques.
Finally, while DRL controllers can be designed to solve specific problems in isolation,
they can also be integrated as parts of larger systems,
e.g.,
along with
classical planning and scheduling modules~\cite{icra:munawar:2018},
which is how we will ultimately envision their common application in the future.

\subsection{Reinforcement Learning for Power-Consumption Optimization}
We now consider the optimization of data-center power consumption as a reinforcement learning problem.
To do so, we constructed a dedicated DRL environment by detailing the definition and computation of the following three items for neural network training: state, action and reward.

\subsubsection{State}
\label{sec:method:state}
The state vector contains the following:
\begin{itemize}
    \item Outdoor air temperature between \SI{-20}{\celsius} and \SI{50}{\celsius}
    \item West Zone air temperature between \SI{-20}{\celsius} and \SI{50}{\celsius}
    \item East Zone air temperature between \SI{-20}{\celsius} and \SI{50}{\celsius}
    \item Total electric demand power between \SI{0}{\watt} and \SI{1}{\giga\watt}
    \item Non-HVAC electric demand power between \SI{0}{\watt} and \SI{1}{\giga\watt}
    \item HVAC electric demand power between \SI{0}{\watt} and \SI{1}{\giga\watt}
\end{itemize}
These ranges are defined as observation\_space when a DRL environment is created and passed to the agent.

\subsubsection{Action}
\label{sec:method:action}
We consider the power-consumption optimization problem as a continuous control task,
i.e., the commands we are interested in can be controlled in continuous domains
(e.g., set a target temperature to between \SI{20}{\celsius} and \SI{30}{\celsius})
rather than discrete
(e.g., flip a switch on or off, set a target temperature to exactly \SI{20}{\celsius},
\SI{21}{\celsius}, etc.).
The action vector contains
\begin{itemize}
    \item West Zone setpoint temperature between \SI{10}{\celsius} and \SI{40}{\celsius}
    \item East Zone setpoint temperature between \SI{10}{\celsius} and \SI{40}{\celsius}
    \item West Zone supply fan air mass flow rate between \SI[per-mode=symbol]{1.75}{\kilo\gram\per\second} and \SI[per-mode=symbol]{7.0}{\kilo\gram\per\second}
    \item East Zone supply fan air mass flow rate between \SI[per-mode=symbol]{1.75}{\kilo\gram\per\second} and \SI[per-mode=symbol]{7.0}{\kilo\gram\per\second}
\end{itemize}
Note that these values are normalized to the range of $[-1.0, 1.0]$ when they are passed from the DRL agent.
Therefore, they must be mapped to actual temperatures by the simulation wrapper described in Section~\ref{sec:environement:wrapper} before sending to the EnergyPlus process.

Given a state vector $\Astatevec_{i}$ computed following Section~\ref{sec:method:state},
an action vector $\Aactionvec_{i}$ is sampled
from the policy network:
$\Aactionvec_{i} \sim \Apolicy_{\Aneuralnetworkparameters}\Aparentheses{\Astatevec_{i}}$.
The $\Aactionvec_i$ is then played
by setting each temperature to the corresponding values and running
the simulation for N steps.
This allows the computation of an updated state vector
$\Astatevec_{i+1}$.
We also define a reward signal $\Arewardval_{i}$
for the purpose of DRL.

\subsubsection{Reward}
\label{sec:method:reward}

We maximize a reward function that consists of a temperature-violation penalty for each zone and the total electrical energy cost.
The reward function for each simulation timestep is divided into two components depending on temperature $r_T$ and power consumption $r_P$:
\begin{equation} 
\label{eq_reward}
r_t = r_T + \lambda_P r_P,
\end{equation}
where $\lambda_P$ is the power-consumption weight and
\begin{equation} 
\nonumber
r_T = -\sum_{i=1}^{z} \exp \left(- \lambda_1 ( T_t^i - T_C^i )^2 \right) - \lambda_2 \sum_{i=1}^{z} \left( [ T_t^i - T_U^i ]_+ + [ T_L^i - T_t^i ]_+ \right), r_P = - P_t,
\end{equation}
where
$z$ is the number of data-center zones,
$T_t^i$ is the temperature of zone $i$ at time $t$,
$T_U^i$ is the desired upper bound temperature for zone $i$,
$T_C^i$ is the desired average temperature for zone $i$,
$T_L^i$ is the desired lower bound temperature for zone $i$,
$P_t$ is the total power consumption at time $t$,
and $\lambda_1$ and $\lambda_2$ are the weights to modify the $r_T$ of Figure~\ref{fig_reward_temperature}.
The $r_T$ consists of Gaussian-shaped and trapezoid penalty parts.
The former gives a maximum reward of 1.0 at the temperature center $T_C^i$, and the reward decreases quickly toward zero as the difference from the center temperature increases.
This makes it difficult to recover to the temperature center once the temperature difference becomes large.
Thus, we added the trapezoid penalty part, which degrades gradually toward $-\inf$.

The values of the reward parameters are chosen, as discussed in Section \ref{sec:results}, depending on criteria such as desired temperatures for human workers and priority with respect to energy minimization

%% file: sections/results.tex
\section{Results}
\label{sec:results}

We discuss the effectiveness of our DRL controller through simulations in EnergyPlus.

\subsection{Setup}

\subsubsection{Building Model}
We used the simulation model ``2ZoneDataCenterHVAC\_wEco\-nomizer.idf'' contained in the standard EnergyPlus 8.8.0 distribution.
It is a data-center model with two thermal zones: West Zone of dimension \SI{232.26}{\metre\squared} and East Zone of dimension \SI{259.08}{\metre\squared}.
The HVAC components include an air economizer, DEC, IEC, Single Speed DX CC, CW CC,
and VAV Fan with a No Reheat Air Terminal Unit.
Zone temperatures are controlled by i) a setpoint manager to control the setpoint temperature of the HVAC components, and ii) thermostat controller set for \SI{23.0}{\celsius} as the cooling setpoint and \SI{20.0}{\celsius} as the heating setpoint.

We added the following modifications to the simulation model in addition to the change in connection to the DRL-based agent described in Section~\ref{sec:environment}.
\begin{itemize}
\item Extend the simulation period from 14 days to 365 days.
\item Change the reference from ``West Zone Inlets'' to ``East Zone Inlets'' in the definition of ``East System Setpoint manager''.
\end{itemize}

When we extended the simulation period, we needed to relax the convergence tolerance in the CalcIndirectResearchSpecialEvapCoolerAdvanced module; otherwise, we would experience severe errors in the EnergyPlus simulation.
We changed the convergence tolerance (TempTol) from 0.01 to 0.02 and the maximum iteration number (MaxIte) from 500 to 1000.

\subsubsection{Weather Data}
The weather information of the data center was simulated using historical events for a year, which are bundled with EnergyPlus 8.8.0 as example weather data.
There are five different weather data files from various locations in the USA, collected and published by the World Meteorological Organization as follows (with average temperatures):

\begin{itemize}
		\item \textbf{CA}: San Francisco Int'l Airport, CA, USA (\SI{13.8}{\celsius})
		\item \textbf{CO}: National Renewable Energy Laboratory at Golden, CO, USA (\SI{9.7}{\celsius)}
		\item \textbf{FL}: Tampa Int'l Airport, FL, USA (\SI{22.3}{\celsius})
		\item \textbf{IL}: Chicago-O'Hare Int'l Airport, IL, USA (\SI{9.9}{\celsius})
		\item \textbf{VA}: Washington Dulles Int'l Airport, VA, USA (\SI{12.6}{\celsius})
\end{itemize}

\subsubsection{Controllers}

We compared simulation results of three different controllers:
\begin{itemize}
\item \textbf{Baseline}: model-based controller built in EnergyPlus.
\item \textbf{TRPO (CA)}: TRPO-based controller trained on CA weather data.
  We chose it because it has the most moderate average temperature among the five weather data files.
  After convergence of the reward value is observed, we alternated between all five training data files, and evaluated how temperature and power controls were performed.
\item \textbf{TRPO (CA-CO-FL)}: TRPO-based controller trained on CA, CO, and FL weather data, which we chose from the five weather data files since they have moderate, coldest, and hottest average temperatures.
  These weather data were switched in every epoch of the simulation process, which is one year long.
  We then alternated between all five training data files.
\end{itemize}

\subsubsection{Hyperparameters for TRPO}

The policy consists of an MLP containing two fully connected layers of size 32 with tanh non-linearities as the activation function.
The hyperparameters for TRPO were set as follows: max\_KL=0.01; timesteps\_per\_batch=16k; cg\_iters=10, cg\_dumping=0.1; gamma=0.99; \seqsplit{vf\_iters=5;} vf\_stepsize=1e-3; lam=0.98. These are default parameters for TRPO except for timesteps\_per\_batch, which was increased from 1k to 16k for stability of the learning process.

\subsubsection{Timesteps}

The timestep parameter for simulation of EnergyPlus was set to four, as defined in the building model.
This means 15 minutes per timestep.
Note that there is another type of timestep used internally in the EnergyPlus simulation process, which is called ``System timesteps''.
Its length varies dynamically ranging from 1 minute/timestep to zone timestep (15 minutes/timestep in this case) to balance simulation precision and speed.
EnergyPlus and the TRPO-based agent communicate with each other in the system timestep frequency.

\subsubsection{Reward-Function Parameters}

We used the following parameters for the reward function~\eqref{eq_reward}:
$T_U^i = \SI{24.0}{\celsius}$,
$T_C^i = \SI{23.5}{\celsius}$,
$T_L^i = \SI{23.0}{\celsius}$,
$\lambda_P = 1 / 100000$,
$\lambda_1 = 0.5$,
and
$\lambda_2 = 0.1$.

\begin{figure}[htpb]
  \begin{minipage}{0.4\textwidth}
    \centering
    \includegraphics[scale=0.55,bb=50 100 410 252]{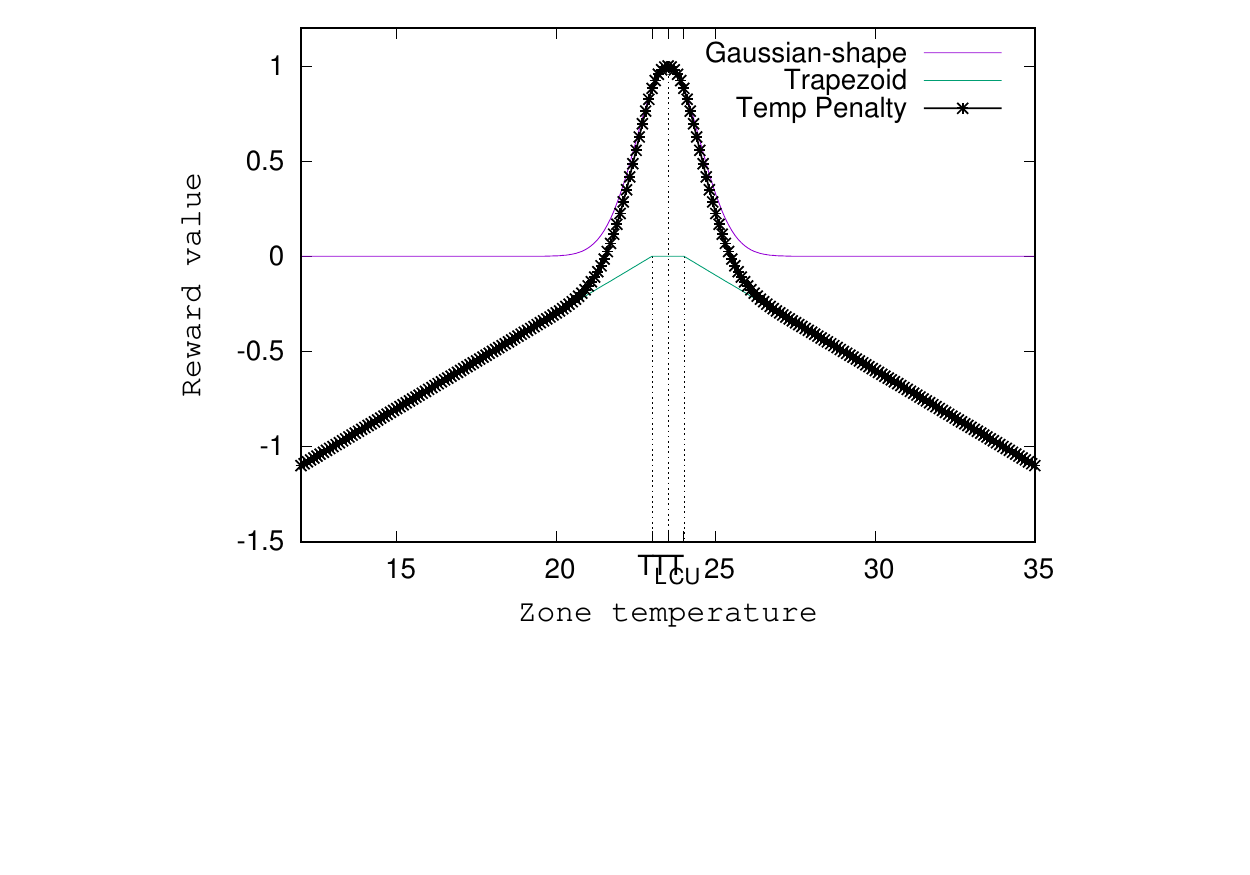}
    \caption{Reward function for temperature $r_T$}
    \label{fig_reward_temperature}
  \end{minipage}
  ~\hfill~
  \begin{minipage}{0.55\textwidth}
    \centering
    \includegraphics[scale=0.55,bb=10 90 410 252]{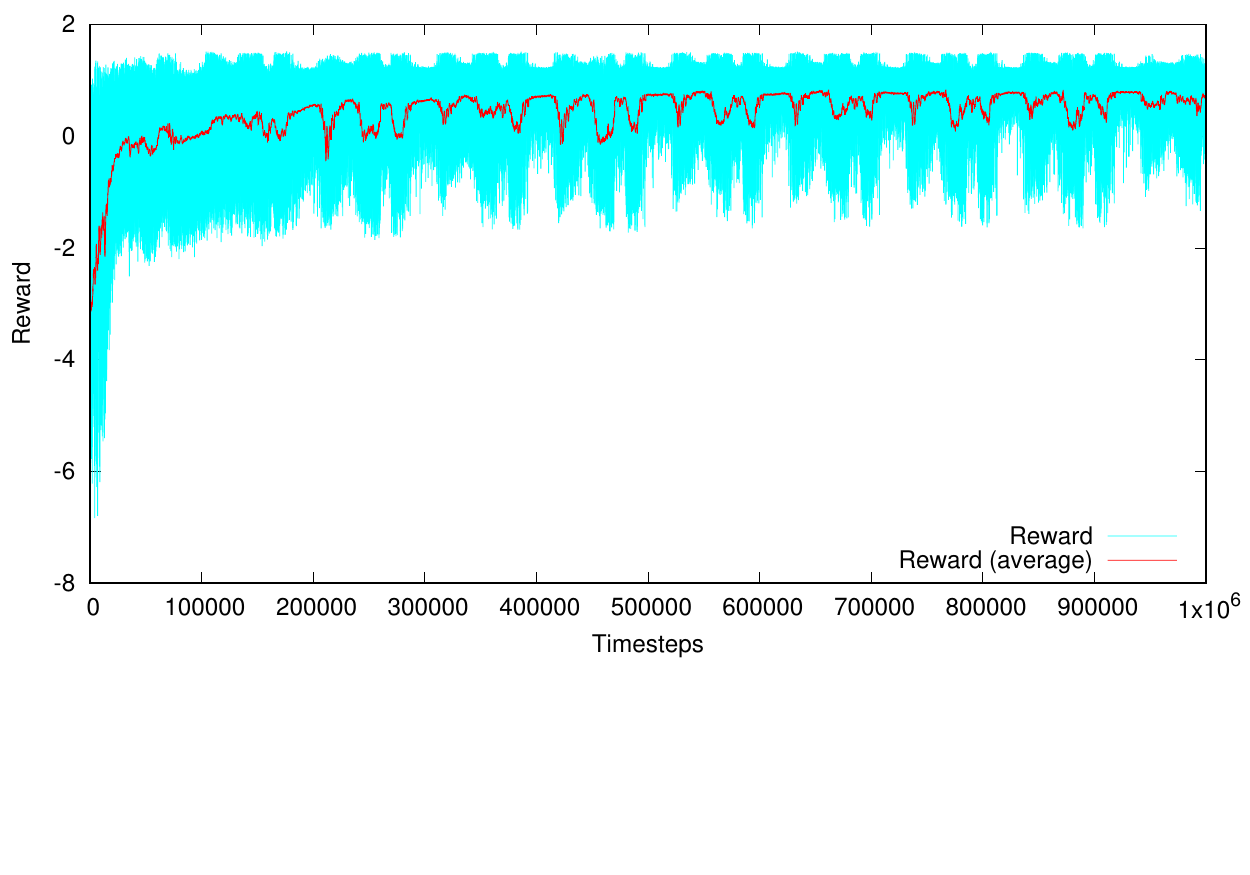}
    \caption{Convergence process of temperature control for TRPO (CA-CO-FL)}
    \label{fig:convergence}
  \end{minipage}
\end{figure}

\subsection{Simulation Results}

Figure~\ref{fig:convergence} shows the convergence process of the average reward of the TRPO algorithm throughout the learning process for training on the CA-CO-FL dataset.
At the beginning, the reward value was low because of the temperature and power penalties.
The average reward increased as the simulation proceeded.
After 200,000 timesteps of simulation, the value was almost stabilized.
We observed some fluttering even after 300,000 steps.
This is because we were alternating between three different weather data files.
Simulation with weather data of a hotter city required higher electrical power for cooling; thus, the reward decreased even though temperature was well controlled.

Table \ref{tab:reward-power-comparison} shows the average reward values and average power consumption with the Baseline algorithm, TRPO trained on CA weather data, and TRPO trained on CA-CO-FL weather data.
In average, the TRPO-based controller achieved 22\% reduction of total electrical demand power.

\begin{table}
  \caption{TRPO was trained on CA or CA-CO-FL then evaluated with five weather data files}
  \label{tab:reward-power-comparison}
  \scalebox{0.75}{
  \begin{tabular}{ l | l l l l l | l l l l l | l l l l l }
    Method (training data) & \multicolumn{5}{c|}{Baseline} & \multicolumn{5}{c|}{TRPO (CA)} & \multicolumn{5}{c}{TRPO (CA-CO-FL)} \\
    Test data              &    CA &    CO &    FL &    IL &    VA &    CA &    CO &    FL &    IL &    VA &    CA &    CO &    FL &    IL &    VA \\
    \hline                                                                                                                                        
    Reward                 &  0.50 &  0.28 &  0.06 &  0.42 &  0.37 &  0.85 &  0.13 &  0.59 &  0.25 &  0.44 &  0.98 &  0.96 &  0.87 &  0.93 &  0.96 \\
    Power consumption (kW) & 117.7 & 122.7 & 140.4 & 124.2 & 125.9 &  97.9 &  99.4 & 113.9 & 100.1 & 100.3 &  98.2 &  94.2 & 108.3 &  94.6 &  96.1 \\
    Average (kW) & \multicolumn{5}{c|}{126.2} & \multicolumn{5}{c|}{102.3 (-18.9\%)} &  \multicolumn{5}{c}{98.3 (-22.1\%)} \\
  \end{tabular}
  }
\end{table}

\begin{figure}
  \begin{tabular}{c}
    \begin{minipage}{0.7\hsize}
      \centering
      \includegraphics[scale=0.6, bb=30 50 390 232]{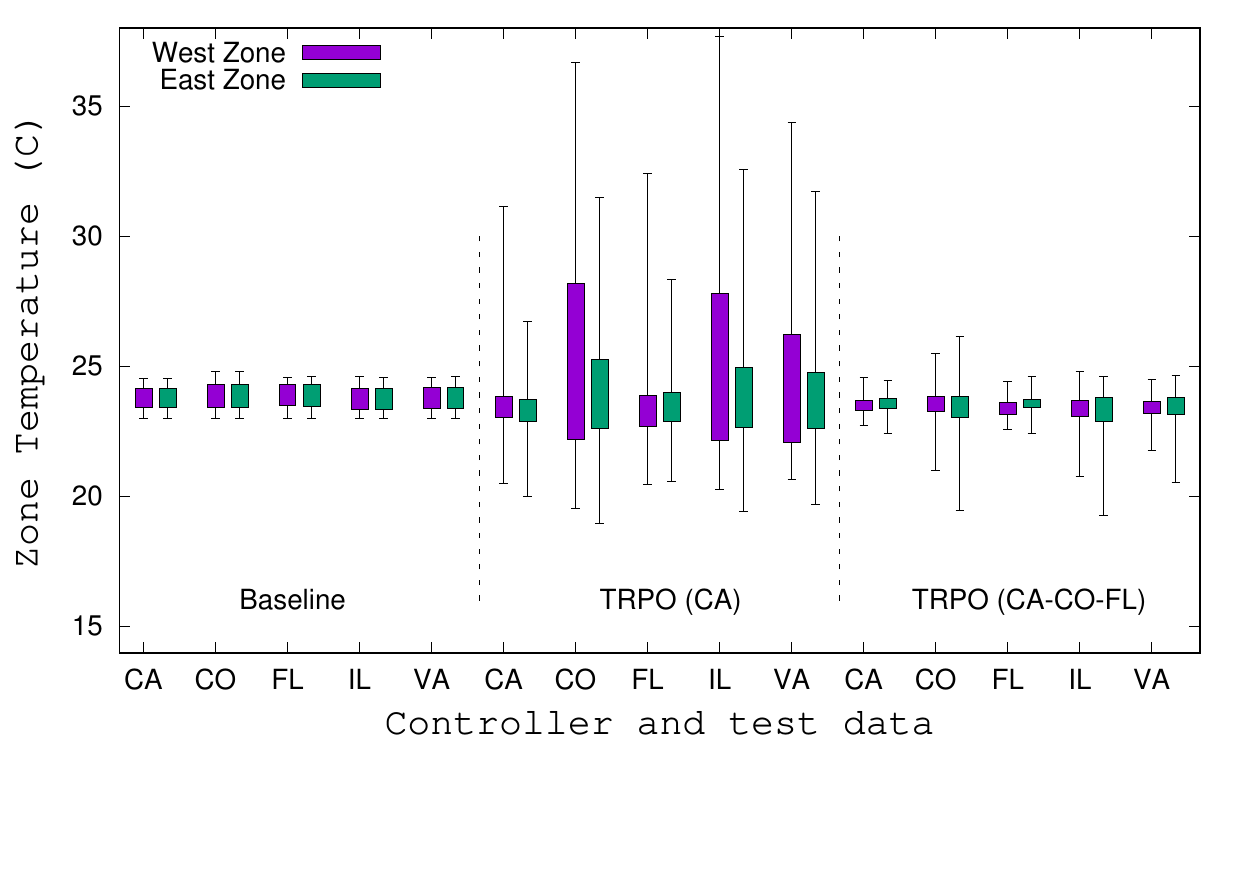}
    \end{minipage}
    \begin{minipage}{0.3\hsize}
      \centering
      \includegraphics[scale=0.4, bb=90 75 450 327]{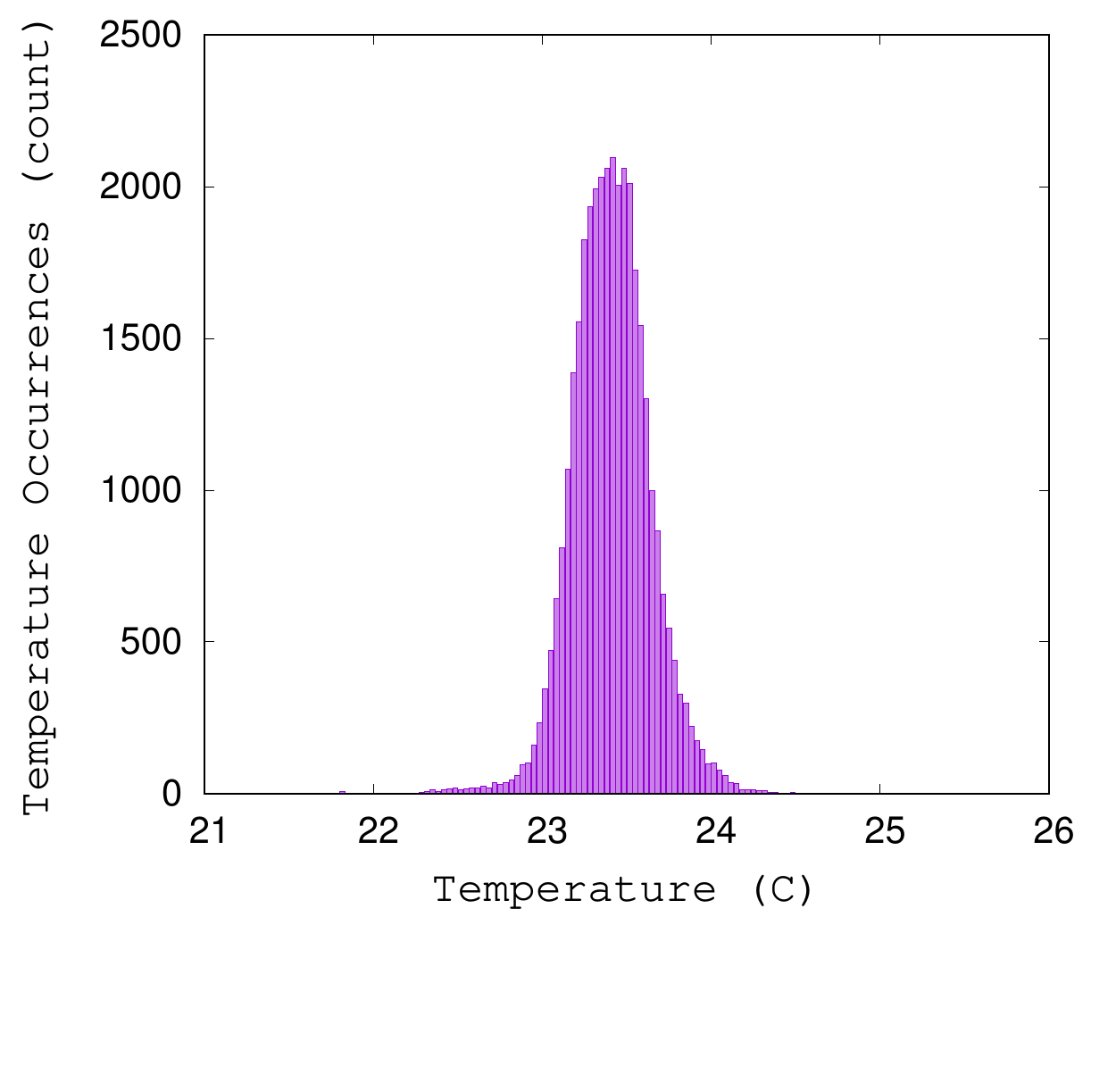}
    \end{minipage}
  \end{tabular}
  \caption{Left: Comparison of controllers regarding zone temperatures. Vertical lines show minimum and maximum temperatures of each zone. Boxes show average temperature $\pm$ standard deviation. Right: distribution of West Zone temperatures for VA dataset with TRPO (CA-CO-FL)}
  \label{fig:temperature_comparison}
\end{figure}

Figure~\ref{fig:temperature_comparison} is a comparison of different controllers regarding zone temperatures.
On the left, the vertical thin lines show maximum and minimum temperatures in a simulation epoch (simulation for 365 days) of each zone.
Boxes on the lines show average temperature $\pm$ standard deviation.

For Baseline, zone temperatures were all distributed within the range of [\SI{22.9}{\celsius},\SI{24.8}{\celsius}], while zone thermostat setpoints were set to \SI{23.0}{\celsius} for cooling and \SI{20.0}{\celsius} for heating.
The average standard deviations of zone temperatures for Baseline was \SI{0.40}{\celsius}.

For TRPO (CA), 300 episodes were executed with training data, then the results were collected for each of the five test data files.
Note that neural networks were updated while these results were collected.
While the standard deviations for CA and FL data were as small as those of Baseline (0.50 on average), the same for CO, IL, and VA were quite large (1.91 in average).
In all cases, the minimum-maximum temperature range was also large.
Just 88.2\% of temperature results fell into the range of [\SI{22.0}{\celsius},\SI{25.0}{\celsius}].
The average standard deviation of these cases was \SI{1.34}{\celsius}.

For TRPO (CA-CO-FL), because we used more training data than for TRPO (CA), the results for IL and VA data were as good as those for the training data.
The average standard deviation was \SI{0.28}{\celsius}, the minimum-maximum temperature ranges were as small as those for the training data,
and 99.6\% of temperature results fell into the range of [\SI{22.0}{\celsius},\SI{25.0}{\celsius}].
The figure on the right shows the distribution of West Zone temperatures for the VA data with TRPO (CA-CO-FL).

%% file: sections/conclusion.tex
\vspace{-0.45\baselineskip}
\section{Conclusion}
\label{sec:conclusion}
Recent advances in artificial intelligence aims to make "intelligent" systems capable of imitating or improving human decision-making systems based on human expertise.
We showed how recent advances in reinforcement learning can be applied to real-world scenarios such as controlling the cooling system of a data center for power-consumption optimization.
Using reinforcement learning techniques, the system is able to outperform the built-in controller by 22\%.
We believe that by releasing the code of our simulation environment, researchers can use our results as a baseline and further improve cooling-system performance by using advancements in reinforcement learning.
In the near future, we plan to use our DRL controller to optimize the energy consumption of an actual IBM data center.